# Multi-source Relations for Contextual Data Mining in Learning Analytics

*Julie Bu Daher, Armelle Brun, Anne Boyer*

*KIWI team, LORIA laboratory*

*University of Lorraine*

*{julie.bu-daher, armelle.brun, anne.boyer}@loria.fr*

**Track**

Academic research: comprehensive evaluations of recent innovations in learning and student analytics approaches.

**Context and Purpose**

The goals of Learning Analytics (LA) are manifold, among which helping students to understand their academic progress and improving their learning process, which are at the core of our work. To reach this goal, LA relies on educational data: students' traces of activities on VLE, or academic, socio-demographic information, information about teachers, pedagogical resources, curricula, etc. The data sources that contain such information are multiple and diverse.

Data mining, specifically pattern mining, aims at extracting valuable and understandable information from large datasets [2]. In our work, we assume that multiple educational data sources form a rich dataset that can result in valuable patterns. Mining such data is thus a promising way to reach the goal of helping students. However, heterogeneity and interdependency within data lead to a high computational complexity.

We thus aim at designing low complex pattern mining algorithms that mine multi-source data, taking into consideration the dependency and heterogeneity among sources. The patterns formed are meaningful and interpretable, they can thus be directly used for students.

**Data and expected output**

As previously mentioned, data comes from various sources. As an example, the *Activity* source is the log file of activities of students on their VLE and contains sequences of quadruplets *<student_id, resource_id, timestamp, action>*. The *Resource* source contains descriptions of pedagogical resources, whether descriptions are structured or not, if not unknown. For example, a resource can have the following attribute values *<Mathematics, Exercise, Difficult, Linear Algebra>*. *Curriculum* describes the different curricula offered by a school and *Students* contains demographic information of students.

Given these 4 sources, a mined pattern could be the following:

*{14-years, Male, Mathematics-grade-9} R-15 R-42 R-Mathematics*

This pattern is made up of two parts. The first one contains information about the student: age (*14-years*) and gender (*Male*) from *Student* and an element about a school program (*Mathematics-grade-9*) from *Curriculum*. The second part is a frequent sequential pattern from *Activity* and *Resource* sources: (*R-15 R-42*) represents a sequence of resources id (from *Activity*) frequently accessed by students; followed by (*R-Mathematics*), which is an attribute from *Resource.* None of the resources accessed following (*R-15 R-42*) are frequent, so they cannot be part of this pattern, but they all share the (*R-Mathematics*) attribute, which is frequent in this context, this attribute is a generalization. The resulting pattern is thus at the same time a specific pattern (as it is valid for *{14-years, Male, Mathematics-grade-9}* students only), and a general pattern.

**Scientific Approach and Challenges**

Pattern mining has mainly focused on mining one data source, but mining multi-source data is becoming an emerging challenge [2]. The literature proposes either to combine data sources together (through the use of the relations between them) and use a single mining process or to mine them independently and then merge the outputs [4], considering the sources of the same importance. In our work, we stress that these relations can be of different types, so they have to be managed differently, resulting in more valuable patterns. For example, a relation can be between two sources (a student *Activity* and a *Curriculum*), or between one source and one element of another source (a resource id from an *Activity* and the description of a *Resource*). Contextual data mining [3] considers two types of data sources: core sources and contextual sources, where a contextual source provides additional information to a core source. A "star-schema" representation was proposed to manage data such that there is one core source and other sources, related/linked to it, are considered as contextual data [1].

The approach we plan to adopt inspires from both previous approaches: a contextual data mining approach that exploits the types of relations/links between sources to guide the mining and obtain valuable patterns, while limiting the complexity.

We propose to consider *Activity* as the core source as it is the one representing the learning process; and other sources as contextual data. Of course, based on the types of links between core and context sources, some of them will be considered as real context (to refine the patterns) and others as additional information (to mine general patterns).

The first challenge is characterizing the data sources and possible relations/links between them. Based on them, the second challenge is defining pattern mining algorithms to automatically manage data sources and their links, and upon needs (specific or general patterns). The numerous possibilities of patterns that can be mined are a major advantage not only to capture the specificities of each student but also to provide personalized help.

**Acknowledgments**

This work has been funded by the PIA2 e-FRAN METAL Project.